# Phase behavior of a deionized binary mixture of charged spheres in the presence of gravity

Nina J. Lorenz*, Hans Joachim Schöpe and Thomas Palberg

Institut für Physik, Johannes Gutenberg Universität Mainz, Staudinger Weg 7, D-55128 Mainz, Germany

Abstract

We report on the phase behavior of an aqueous binary charged sphere suspension under exhaustively deionized conditions as a function of number fraction of small particles $p$ and total number density $n$. The mixture of size ratio $\Gamma = 0.557$ displays a complex phase diagram. Formation of bcc crystals with no compositional order dominates. We observe a region of drastically decreased crystal stability at $0.55 < p < 0.95$ with the minimum located at $p = 0.8 \pm 0.05$ at densities above $n = 26 \mu m^{-3}$. A peaked region of enhanced stability is observed at $p = 0.4$. Further light scattering experiments were conducted to characterize the crystallization time scales, the density profiles and the composition of formed phases. For $0.82 > p > 0.95$ crystal formation is partially assisted by gravity, i.e. gravitational separation of the two species precedes crystal formation samples in the coexistence range. In the composition range corresponding to the decreased crystal stability only lower bounds of the freezing and melting line are obtained, but the general shape of the phase diagram is retained. At $p = 0.93$ and $n = 43 \mu m^{-3}$ two different crystalline phases coexist in the bulk, while at $p = 0.4$ additional Bragg peaks appear in the static light scattering experiments. This strongly suggests that we observe a eutectic in the region of decreased stability, while the enhanced stability at $p = 0.4$ seems to correlate with compound formation.

# 1. Introduction

Sterically stabilized and/or charge stabilized one component colloidal dispersions have for long been regarded an excellent model system for questions of statistical mechanics and solid state physics [1, 2]. The steepness of the stabilizing repulsion is conveniently varied by experimental means between the theoretical limits of the one component plasma and the hard sphere case [3]. A liquid-gas phase separation is absent as the interaction is purely repulsive, but at constant volume colloidal crystals are formed from colloidal fluids once the strength and range of Coulomb repulsion is sufficiently large or once a critical hard sphere packing fraction is exceeded [4, 5, 6]. In charged sphere systems at low particle concentrations body centered cubic (bcc) crystals are formed while at higher concentrations and in hard sphere systems close packed crystal structures prevail [7, 8]. Investigations meanwhile have covered also non-equilibrium situations like crystallization kinetics and the approach of the glass transition [9] or the influence of external fields [10].

Recently, interest shifted towards mixtures, where several additional parameters become important for the phase behavior and system properties. Most important are the size ratio $\Gamma = a_S / a_L$ (where $a$ denotes the particle radius and the suffixes S and L refer to the small and large component, respectively), the charge ratio $\Lambda = Z_{eff,S} / Z_{eff,L}$ (where $Z_{eff}$ denotes the effective charge of each pure species) and the composition $p = n_S / (n_S + n_L)$ (where $n$ denotes the particle number density). Three types of mixtures have attracted particular interest. First, oppositely charged spheres, where Coulomb attraction dominates and a wealth of salt structures are found [11, 12]. Second, highly asymmetric mixtures, neutral or of like charge, where entropically driven phase separation [13, 14] dominates both phase behavior and transition kinetics. Here stable gas, liquid and crystal phases, but also metastable gels, clusters and an attractive glass are observed [15, 16, 17, 18, 19].

Thirdly, less asymmetric mixtures, where – depending on the details of the interactions – crystallization into pure component crystals, solid solutions or stoichiometric compounds as well as the formation of amorphous states are observed. For instance, hard sphere or strongly screened charged sphere mixtures form either stoichiometric compounds or amorphous states [20, 21, 22, 23, 24, 25, 26, 27, 28]. Early theoretical work on hard spheres suggested the formation of zero miscibility eutectic phase diagrams. Later work revealed a low, but finite miscibility leading to the formation of compounds at certain size ratios and compositions with peritectic and other types of more complicated phase behavior [29]. Simple, eutectic phase behavior is not any longer expected to dominate and in fact has not yet been seen in experiments on hard spheres. Substitutionally ordered solid solutions with spindle type or azeotropic phase diagram are restricted to large size ratios $\Gamma > 0.8$ [30, 31]. The experimentally often observed vitrification is aided by the closeness of the kinetic glass transition, but recent calculations show, that it is also connected to the unavoidable intrinsic polydispersity [32].

By contrast, for charged sphere mixtures at low salt conditions the repulsion is soft, in particular under low salt conditions. Therefore, charged sphere mixtures are much less susceptible to size mismatch even off stoichiometric composition. Generally they are well miscible in both fluid and crystalline state. With increasing $n$ one observes stable fluids, substitutionally ordered solid solutions of bcc structure and finally of close packed crystal structure [33, 1, 34, 35, 36, 37]. In a few cases compound formation was observed at densities considerably larger than the melting density [35, 38, 39, 40]. Concerning the phase diagram type, Meller and Stavans [34] proposed that as a function of composition the shape of the phase boundaries in a $1/n$-$p$ representation varies with increased asymmetry from spindle type over azeotropic to eutectic. This was connected to an increasingly lowered miscibility of the components in the solid state similar to the reasoning in hard spheres, atomic and molecular substances. The conjecture was supported by experimental phase diagrams showing a monotonically increasing freezing density at $\Gamma = 0.87$ and a peaked freezing at $\Gamma = 0.78$. At $\Gamma = 0.54$ two crystalline regions (of undetermined composition) were separated by a fluid region. Unfortunately, the lower azeotropic or eutectic point was not reached and neither formation of a substitutionally ordered alloy nor the coexistence of two crystal species could be demonstrated. Rather, at large concentrations an amorphous solid bordered the fluid phase. In a study on exhaustively deionized charged sphere mixtures Okubo and Fujita [35] reported a transition from spindle type to azeotropic phase behavior with coexisting fluids and solid solutions for size ratios decreasing from $\Gamma = 0.93$ to $0.77$. In our own group we could show that systematic variation of composition for spindle type phase behavior was accompanied by a number weighted variation of other properties like conductivity, elasticity or solidification kinetics [36, 37], a concept originally introduced by Lindsay and Chaikin [33]. Finally, Kaplan et al. have shown that in charged but strongly screened mixtures of moderate asymmetry ($\Gamma \geq 0.2$) instead of bulk phase separation a surface phase separation at the container walls may occur, including the possibility of surface crystallization of the larger component [41]. Thus the cases of rather large and of very small size asymmetry have already received some attention, while the even more interesting intermediate region of parameter space is much less investigated.

The present paper goes beyond previous work in several aspects. First we study a mixture at $\Gamma = 0.557$ ($\Lambda = 0.568$) under thoroughly deionized conditions. Here the soft and long ranged repulsion increases the mobility and prevents dynamic arrest in an amorphous state. Therefore we are able to assess the fluid-crystal phase boundary over the full range of compositions. Moreover, for the first time also the fluid-solid coexistence region is mapped out in some detail. Second, we pay particular attention to the effects gravity. For some compositions the separation of species by sedimentation occurs on time scales comparable to or even smaller than those of solidification. This influence is characterized in detail and discussed with respect to possible consequences for the location of the melting and the overall phase diagram type. We show, that for some compositions the location of the phase boundaries are imprecise, but the type of phase diagram is retained. Finally, the chosen asymmetry to significantly lowers the miscibility in the solid state as compared to the isotropic fluid except for the case of compound formation. Our detailed study reveals a region of drastically decreased crystal stability, also observed in previous work. Here, however we are able to provide further evidence of eutectic behavior from microscopic studies of the sample morphologies. In addition, an enhanced crystal stability is observed at $p = 0.4$, which seems correlated with compound formation at high particle concentrations. The shape of the observed phase diagram thus is considerably more complex than previously reported ones.

The rest of the paper is organized as follows. We first introduce our samples and their conditioning and shortly sketch the employed methods. We then give some general observations and discuss several aspects of the crystal formation process in some detail Next the phase behavior is presented and discussed in connection with crystallographic observations by microscopy of thin cell samples. In the discussion we compare our findings to previous results and to expectations from theory. We close by summarizing and identifying some interesting open questions.

## 2. Experimental

### 2.1    System and system conditioning

Two industrial species of poly-*n*-butylacrylamide-polystyrene copolymer spheres stabilized by Sulphate surface groups were employed, which were a kind gift of BASF, Ludwigshafen. Prior to the present experiments both species were extensively characterized and Tab. I compiles the most important properties. As can be seen the size ratio is $\Gamma = 0.557$ (as obtained from ultracentrifugation) and the charge ratio is $\Lambda = 0.568$ (as obtained from elasticity measurements, see below). The particle mass density is 1.05g/cm. From theoretical estimates are available for dilute and non-interacting one component samples we expect an influence of gravity on our investigations. Under gravity an exponential density profile corresponding to the barometric height formula is expected [42]. Its decay constant, the sedimentation length is given as $\ell = (k_B T)/(g\Delta\rho V_P)$, where $g$ = 9.81ms$^{-2}$ is the gravitational acceleration, $\Delta\rho$ = 50kg/m$^{-3}$ is the mass density difference between our particles and the solvent and $V_P$ is the particle volume. This length amounts to 8.8mm for our large particles and to 51mm for the small. Using Stokes law, the large spheres would settle approximately 3.2 times faster than the small. Thus, in principle, not only sedimentation, but also segregation may occur in our experiments. For strongly interacting suspensions the situation is much more complex, even for one component systems [43, 44]. Predictions for exhaustively deionized mixtures are not yet available. We therefore have to estimate the influence of gravity on the location of phase boundaries and the phase diagram type from additional experiments.

Preparation of exhaustively deionized mixtures proceeded as follows. From the batches supplied we first prepared two stock suspensions of approximate particle density $n \approx 35\mu m^{-3}$ by dilution with doubly distilled water (Purelab Classic DI, ELGA, UK). To these well-rinsed mixed bed ion exchange resin (Amberlite UP 604, Rohm & Haas, France) was added and the suspensions were left to stand for a week under occasional stirring. They were then filtered using a syringe filter of a pore size of 1.2µm to remove dust and coagulate regularly occurring upon first contact with IEX. The suspensions were then stored over a fresh portion of IEX for several months under occasional mild stirring. Both pure species soon crystallized as the salt concentration was reduced by the action of the IEX. Both formed body centred cubic (bcc) crystals as identified from their static light scattering patterns. Static light scattering further yielded the exact number densities which were adjusted by further dilution to obtain identical $n$ for both species in the stock suspensions.

Samples of desired composition $p = n_{68} / (n_{68} + n_{122})$ (molar or number fraction) were prepared by mixing small amounts of the stock suspensions with distilled water and deionized using a batch procedure, following [35] with an improved protocol. From each mixture series of samples with different particle concentrations were prepared in 2ml cylindrical vials (Supelco, Bellefonte, PA, USA). The uncertainty $\Delta n/n$ from this dilution process is smaller than 10%. Where necessary, $n$ was determined with an uncertainty better than 2% by static light scattering. A small amount of mixed bed ion exchange resin (Amberlite, Rohm& Haas, France) was introduced and the vials were sealed with an air tight Teflon® septum screw cap, carefully avoiding gas bubbles. Samples were placed cap down on a shelf to avoid uncontrolled salt gradients leading to accumulation at the IEX [45]. Under gentle daily stirring they were exhaustively deionized over several weeks. The theoretical limit of the residual ion concentration is given from the ion product of water: $[H^+][OH^-] = 10^{-14}$ mol$^2$ l$^{-2}$, where the proton concentration is set by the amount of counter-ions released by the particles: $[H^+] = n\, Z_\sigma / (1000\, N_A)$ where $N_A$ is Avogadro´s number and $Z_\sigma$ is the electrokinetic charge determined from conductivity [46], which corresponding to the number of freely moving counter-ions. For the typical cases investigated here this amounts to $c_B = [OH^-] \leq 10^{-8}$ mol l$^{-1}$. The residual ion concentration cannot be directly measured in batch conditioning. A very sensitive marker of exhaustive deionization, however, is given by the size of the crystallites, reappearing after shaking the sample. As detailed elsewhere [47], the crystallite size strongly depends on the deionization state via the homogeneous nucleation rate density. The crystallite size decreases with decreasing residual ionic contaminations. A constant small crystallite size thus indicates completion of the deionization process [48]. After two months the samples were homogenized for a last time and left to stand undisturbed on a vibration free shelf. This defined $t = 0$ for the phase diagram determination. The melting and freezing densities obtained thereafter for the pure samples are shown in Tab. I. These are somewhat below those obtained previously using a different deionization procedure with slightly larger residual stray ion concentration. Batch conditioning is thus rather time consuming, but enables exhaustive deionization and facilitates to prepare a large number of samples from a given limited stock supply.

Batch conditioning was also used for the microscopic experiments employing a commercial Microlife[R] cell [Hecht, Germany]. In these, two reservoirs are connected by a thin slit. Each reservoirs contains approximately 1.5ml of suspension, the actual parallel plate measuring chamber spans 47mm in x direction, has a width y = 7.5mm and a height z = 250µm. Both reservoirs contained about 1ml of IEX. Also this cell was tightly sealed by screw caps, but unfortunately the glue used in fabrication is not completely $CO_2$ tight. Hence the residual ion concentration is much larger than for the cylindrical vials. These cells cannot be employed for determination of the phase behavior, but here they allow for detailed morphological studies.

| Sample Batch number | $2a$ / nm | $\ell$ / mm | $v_{Stokes}$ / nm/min | N | $Z_\sigma$ | $Z_{eff}$ | $n_F/\mu m^{-3}$ | $n_M/\mu m^{-3}$ |
|---|---|---|---|---|---|---|---|---|
| PnBAPS68 BASF2168/7387 | 68±3 | 51 | 33.9 | 1435±50 | 450±16 | 331±3 | 1.9±0.3 | 2.6±0.3 |
| PnBAPS122 BASF2035/7348 | 122±3 | 8.8 | 109.2 | 3800±100 | 743±40 | 582±18 | 0.06±0.05 | 0.08±0.05 |

Table I: Properties of the employed charged sphere suspensions. $2a$: diameter and standard deviation from ultracentrifugation; $\ell$ sedimentation length; $v_{Stokes}$: Stokes sedimentation velocity; $N$: Sulfate surface group number from titration, $Z_\sigma$: electrokinetic charge from conductivity (corresponding to the number of uncondensed counter-ions [46]); $Z_{eff}$ effective charge from shear-modulus measurements; $n_F$ and $n_M$ denote the freezing and melting number

densities in the exhaustively deionized state obtained for batch conditioning in cylindrical vials.

*2.2 Experimental methods*

The phase behavior after exhaustive deionization was obtained from visual inspection. Formation of crystallites is readily seen from the appearance of brightly iridescent Bragg reflections under white light illumination. Samples were imaged using a digital camera equipped with a macro lens. Samples in the flat cells were also observed in reflection mode using a diffuse white light illumination combined with a slightly divergent white light illumination mounted on an adjustable holder. The latter allowed tuning of the angle of incidence to collect the Bragg reflections of suitably oriented bulk crystallites (c.f. Figs. 8a and 9a) [49]. Thin wall crystals of only few particle layers, which are oriented with their principal plane parallel to the cell wall cannot be observed this way. Here, the illumination was performed under an angle of 75-85° with respect to the cell normal. The camera observed the sample under a similar angle from the opposite side. This scattering geometry corresponds to grazing incidence x-ray scattering. For few layer crystals, the observed Bragg reflections do not depend on the orientation of the crystals in the cell wall plane, as the scattering originates from the layer spacing only [50] (c.f. Fig. 8c). Samples were also observed by polarization microscopy. Here parallel white light was used for illumination under normal incidence with respect to the cell wall. The sample was observed in transmission between crossed polarizers. This technique exploits the fact that the Bragg scattered light is polarized. Any scattering crystal therefore not only weakens the transmitted light of the corresponding wave length but moreover rotates its polarization direction. Fluid regions and crystals, which are not scattering under normal incidence, do not influence the transmitting light. In this mode Bragg reflecting crystals appear brightly colored on a black background (c.f. Figs. 8d and 9 b-d).

Structure determination and measurements of the particle number density was performed by standard static light scattering employing a Debye-Scherrer geometry. The scattering vector $q = (4\pi\nu/\lambda)\sin(\Theta/2)$, where $\nu$ is the index of refraction of the suspension, $\lambda$ is the vacuum wave length of the laser light, and $\Theta$ is the scattering angle. From the sequence of peak positions all samples (except for few samples at $p = 0.4$) were identified to crystallize in a body centered cubic structure. The number density $n$ is related to $q$ at the position of the Miller indexed peaks of the static structure factor as:

$$n_{bcc} = 2\left(\frac{2\nu}{\lambda}\frac{1}{\sqrt{h^2+k^2+l^2}}\right)^3 \sin^3\left(\frac{\Theta}{2}\right) \qquad (1)$$

This expression was also used to evaluate the position of the principle peak of the fluid structure factor, as close to the phase transition the fluid shows a bcc short range order [51, 52]. Uncertainties in derived $n$ are largest close to the phase boundary due to elastic straining of crystals but decrease to below 2% for larger $n$. Vertical shift of the cell allows determining the structure and density as a function of height.

Dynamic light scattering using a standard homodyne photon correlation set-up with an ALV-5000 correlator was used to determine the apparent radii $a_{app}$ of dilute, non-interacting mixtures. To this end samples of different $p$ were diluted to $n = 0.025\mu m^{-3}$ and left at contact to air to obtain $CO_2$ saturated suspensions. Under these conditions the static structure factor is equal to unity and particles may be assumed to be non-interacting. The measured normalized intensity autocorrelation function is evaluated for its first cumulant. I.e. it is approximated as $g(q,t) \approx \exp(-t/\tau)$ with $\tau = 1/D_{app} q^2$, where the apparent diffusion coefficient $D_{app} = k_B T / 6\pi\eta a_{app}$ relates $a_{app}$ to the thermal energy $k_B T$ and the viscosity $\eta$. Plotting $a_{app}$ versus $p$ allowed to construct a calibration curve from a least square fit of a function $y = \exp(a+b/x+c)$ to the data points. This function is of no physical meaning, but as seen below in Fig. 4, gives a good description of the data. The calibration curve was used to determine the composition of coexisting crystals and fluid regions. To that end small amounts of suspension were drawn from the regions of interest, conditioned as sketched above and their apparent radii measured and compared to the calibration curve.

Measurements of the shear modulus employed the non-invasive optical technique of torsional resonance detection [53]. In short, the sample is set into low amplitude torsional oscillation about its vertical axis. Using a position sensitive detector and lock-in technique the periodic response of the Bragg angle to the periodic crystal straining is recorded and analyzed for the resonance spectrum. At known geometry the spectrum is evaluated for the shear modulus [54]. For several samples of the present study we determine the shear modulus as a function of time, to obtain an estimate of the crystallization kinetics. Shear moduli of equilibrated samples were further evaluated for effective charges [52, 55].

## 3. Results

### 3.1   General observations

After exhaustive deionization the samples were shaken up again and left to crystallize. The general crystallization behavior depended strongly on both composition and particle concentration. Samples with particle density below freezing ($n < n_M$) stayed fluid. For samples with particle density above melting ($n \geq n_M$) complete crystallization occurred within seconds to a few days. For samples at coexistence ($n_F \geq n > n_M$) we could discriminate three different scenarios. The pure samples and all mixtures with $p \leq 0.7$ complete crystallization before the crystals settle subsequently. After some weeks a stationary state was obtained. For $0.82 \leq p < 0.93$ the mixtures stayed fluid over extended times, before crystals formed and stayed in the top part of the cell. These crystals are henceforth termed head crystals. Fig. 1 displays a set of images taken on representative samples. One clearly discriminates crystalline regions from their Bragg reflections. Examples in Fig. 1 include fluid state (Fig. 1a), settled crystals at coexistence (Fig. 1b), head crystals at coexistence (Fig. 1c, 1d). Head crystals also appeared in large-volume vials of $0.82 \leq p < 0.93$, from which sufficient amounts of head crystals and of the bottom fluid could be isolated and further analyzed after repeated deionization (see below). Probes taken from the head crystals fully solidified (Fig. 1f) while probes taken from the lower part did not (Fig. 1g). This shows that crystal formation here is non congruent, i.e. the initial melt and the final crystal differ in composition. In the next chapter we shall present additional measurements to find out, whether this is driven by thermodynamics or due to an influence of gravity.

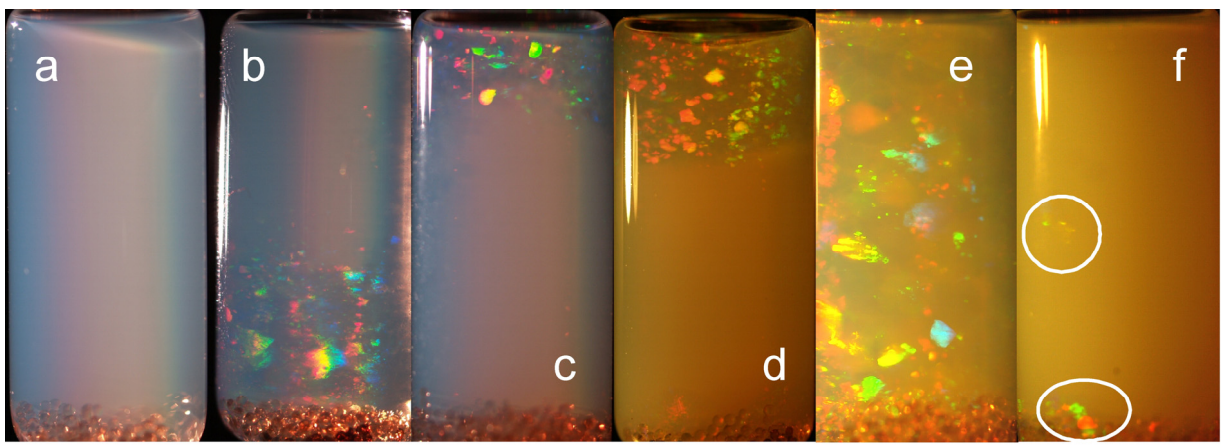

Fig. 1: (color online) Selected examples of phases and their distribution in 2ml vials after exhaustive deionisation one to three months after last shaking. a) $p = 0.9$, $n = 3\mu m^{-3}$ a homogeneous rainbow colored appearance indicates a fluid ordered phase. b) $p = 0.2$, $n = 1.12\mu m^{-3}$ crystalline phase with Bragg reflection at the bottom. c) $p = 0.9$, $n = 6.63\mu m^{-3}$ and d) $p = 0.9$, $n = 12\mu m^{-3}$ head crystal formation at the cell top. e) crystalline phase grown from a head crystal region of a larger vial at $p = 0.9$, $n = 12\mu m^{-3}$. f) mostly fluid phase with few isolated crystals (encircled) from a bottom region of a larger vial at $p = 0.9$, $n = 12\mu m^{-3}$.

*3.2 Sedimentation*

Between one and two months after last shaking both sedimentation processes and structure conversion processes had ceased and the samples had become stationary. We measured the static light scattering patterns at different heights above the IEX of thorough deionized samples prepared at their phase transition points in a state of fluid-crystalline coexistence. In Fig. 2a-c we display three examples obtained for samples with settled crystals and with head crystals. In Fig. 3a-c we plot the corresponding number densities as a function of total cell height.

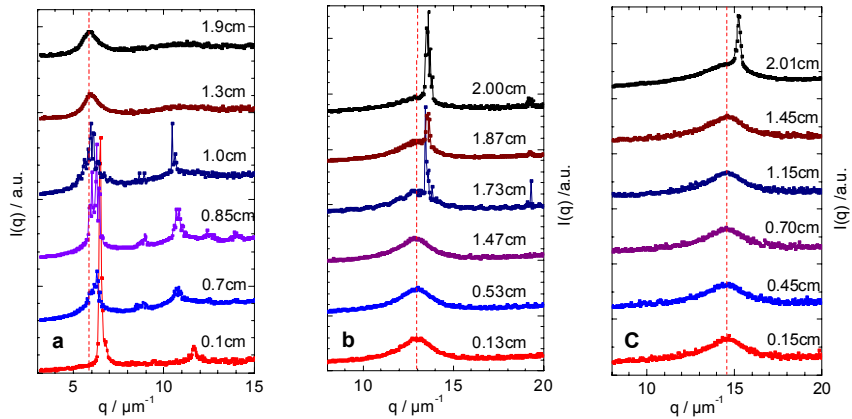

Fig. 2: (color online) Height dependent static light scattering patterns of samples at coexistence measured at the heights above IEX as indicated. a) initial sample parameters: $p = 0.15$, $n = 0.7 \mu m^{-3}$, sample situated at melting number density. At lower height the sample is bcc crystalline whereas the upper part shows fluid structure; Note that the positions of the fluid peak and the first Bragg peak at mid cell height nearly coincide. b) initial sample parameters: $p = 0.9$, $n = 6.2 \mu m^{-3}$. At lower height the sample shows fluid structure whereas the upper part shows a superposition of a fluid scattering signal and crystalline Bragg peaks. Here the positions are shifted with respect to each other. c) initial sample parameters: $p = 0.9$, $n = 9.2 \mu m^{-3}$. Overall appearance as before, but at low heights the position of the fluid maximum shifts towards larger $q$.

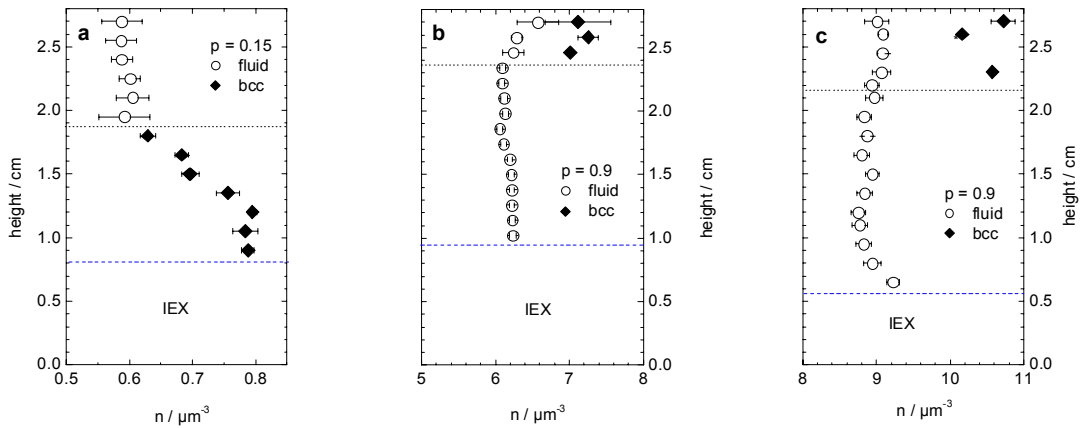

Fig. 3: (color online) number densities as obtained from static light scattering as a function of sample height. Samples as in Fig. 2a-c. The dotted horizontal line marks boundary of the crystalline regions. a) initial sample parameters: $p = 0.15$, $n = 0.7 \mu m^{-3}$, a decreased density is observed in the upper, fluid part, in the crystalline part the density increases above the initial value. Fluid and crystal densities coincide at the boundary. b) initial sample parameters: $p = 0.9$, $n = (6.2 \pm 0.1) \mu m^{-3}$, the fluid region appears to be slightly depleted of particles, an increased number density is observed in the head crystals. c) initial sample parameters: $p = 0.9$, $n = 9.2 \mu m^{-3}$, the fluid region appears to be slightly depleted of particles. The head crystal shows an enhanced density. With increasing height the fluid density first decreases then slowly increases.

The sample with settled crystals shows a decreased density in the upper, fluid part. In the crystalline part the density is increased above the initial value. Fluid and crystal densities coincide at the boundary, which is indicated by the horizontal dotted line in Fig. 3. This sample is representative for the pure samples and for mixtures of $p < 0.7$ and $p > 0.95$. It shows gravitational settling and compression of the crystalline region, which is of body centered cubic structure. Qualitatively, the shape of the density profile compares well to examples from literature [44].

The head crystal forming samples show fluid structure at low height, while in the head crystal region crystalline Bragg peaks are superimposed to the right of the fluid maximum. Here crystals and fluid coexist. In Figs. 3b and c the fluid region is depleted of particles, while the number density of the head crystals increased above the initial density. From mid cell height the fluid phase number density increases with height. Both samples are representative for the head crystal forming samples found to the right of the minimum in crystal stability in the *n-p* phase diagram. For the sample in Figs. 2c and 3c an additional feature in the lower cell part became visible, as this sample contained a smaller amount of IEX. In the lowest part of the cell the fluid phase number density increases pronouncedly as the cell bottom is approached.

Figs. 2 and 3 show characteristic density profiles obtained in the coexisting samples after equilibration. But also the crystallization process itself was observed to be qualitatively different. In the first case, represented by *p* = 0.15, crystals formed at all heights in the cell after short times and subsequently settled. After settling the profiles were stationary. In the case of $0.82 \leq p < 0.93$, such as the samples b and c, crystals form merely in the top region usually within a week and the volume occupied by crystals slightly increases during the second week. Visual inspection of the IEX region showed an enhanced turbidity at the very cell bottom.

*3.3    Composition*

The composition of head crystal and of the fluid region close to the IEX were estimated for a sample of *p* = 0.9 and *n* = 12μm$^{-3}$ (c.f. Fig. 1d) contained in a 100ml vial, where also head crystal formation had occurred. A small amount of suspension isolated from the head crystal region showed complete crystallization after batch deionization (c.f. Fig. 1e), while another taken close to the IEX developed only few crystals adjacent to either wall or the IEX (c.f. Fig.

1f). Both samples were diluted to $n = 0.025\,\mu m^{-3}$ and left open to saturate with $CO_2$. The apparent radii under these non-interacting conditions are compared in Fig. 4 to the calibration curve obtained before (see experimental section). For the diluted head crystal and bottom phase we obtained apparent radii of $a_{app} = (60.6\pm4.8)$nm and $a_{app} = (68.7\pm1.6)$nm corresponding to $p = 0.93\pm0.05$ and $p = 0.79\pm0.05$, respectively. These two values are significantly different, and we conclude, that the head crystal is enriched in small spheres.

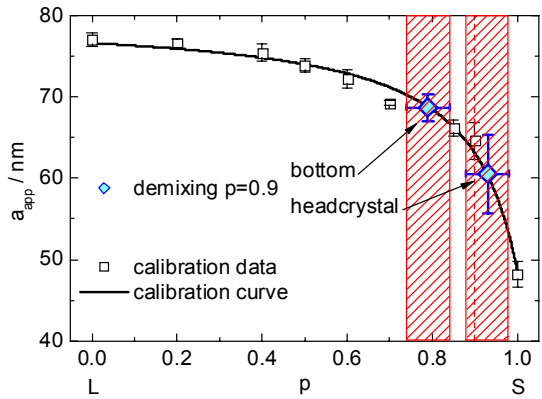

Fig. 4: (color online) Apparent radii of dilute, non-interacting mixtures as a function of composition. Black squares: radii obtained for mixtures of known composition. Solid line: fit of a function $y = \exp(a+b/(x+c))$ to parameterize the data. Open diamonds: radii measured in diluted samples taken from different regions of a phase separated mixture at initial $p = 0.9$ (dashed vertical line) and $n = 12\,\mu m^{-3}$ as indicated. The vertical error bars given correspond to the combined uncertainties of the original data and the fit. They translate to two significantly distinct composition ranges for the head crystal and the supra-natant (hatched areas).

### 3.4     Model for head crystal formation

The two different scenarios for samples in the state of fluid-crystalline coexistence are correlated with the phase diagram shape (see below). Settling crystals are observed on the left side of the crystal phase's stability minimum and head crystals on the right side. The results of our analysis may be combined with this observation, to explain the formation and staying of head crystals with their increased *n* and *p*. During the first one or two weeks a gravity driven vertical demixing of the suspension occurs. The large particles form zones of increased density at the very cell bottom, while the supernatant is at the same time enriched with small particles. Once the composition in the upper part has shifted to sufficiently large *p,* crystal formation occurs, as the number density at the cell top is above $n_F$ of pure small spheres.

Note, that also in the top region of the cell, the number densities in the coexisting fluid and the crystalline phase are slightly different. As the crystallization occurs after gravity driven segregation, this additional small segregation has to be of thermodynamic origin. Fluid and solid coexist at the compositions and densities corresponding to the liquidus and solidus conditions. The same applies to the suspension drawn from the lower part of the large vial at *n* = 12μm$^{-3}$ and *p* = 0.9 shown in Fig. 1f.

For the remaining fluid phase in the lower part the number density is decreased and at the same time enriched with large particles, and its composition is shifted towards the minimum of crystal stability. Thus no crystals may form in the lower parts of the cell. Finally, the mass density of the suspension is given by $\rho = \Phi\rho_{Part} + (1-\Phi)\rho_{Solvent}$ with the volume fraction $\Phi = (4\pi/3)(n_S a_S^3 + n_L a_L^3)$ and $n_S + n_L = n$. Using the numbers determined for the densities and the composition in head crystals and in the fluid part, one recognizes that the mass density of the head crystal is smaller than that of the supra-natant. Hence the crystals stay at the cell top after their formation.

### 3.5  *Bulk crystallization kinetics*

Above the melting densities all samples crystallized on time scales of minutes to days well before noticeable gravitaional settling occurred. To obtain an estimate of the crystallization kinetics and of the composition we measured the shear moduli of selected samples as a function of time (Fig. 6). As seen before for samples at coexistence, also for densities above $n_M(p)$ the solidification kinetics of mixtures with $0.82 \leq p < 0.93$ are considerably slower than for all other compositions. Fig. 5 shows a representative series of photografic images taken on a sample at $p = 0.9$ and $n = 32\mu m^{-3}$. After a few hours, crystallites formed all over the sample but it took at least a day until any shear rigidity could be obtained. Furthermore, the shear modulus G first increased considerably in time till it reached a plateau value after one or two more days, a behaviour consistent with a percolated network of crystallites forming and sealing itself, visible in the photographic images of Fig.5. The saturation values are larger than those of the pure small sphere samples of like number density. Samples of like composition but lower $n$ showed a similar behaviour but on even longer time scales.

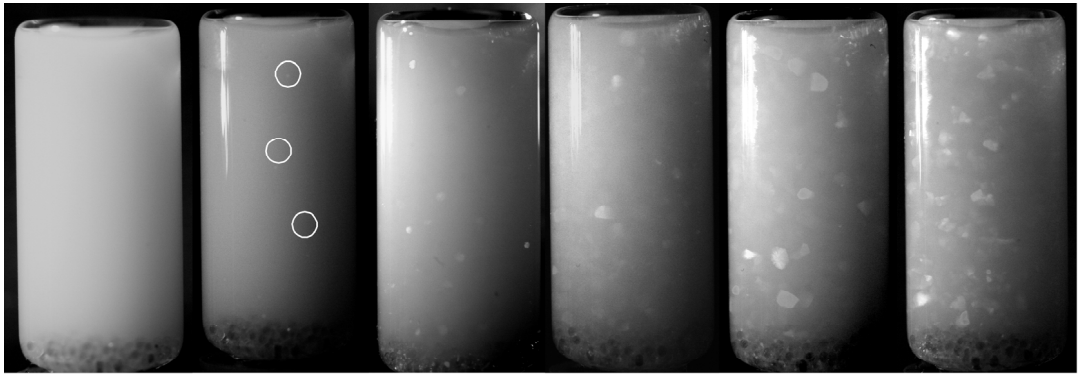

Fig. 5: Series of photographic images taken on an exhaustively deionized sample of $p = 0.9$ and $n = 32\mu m^{-3}$ as a function of time. From left to right, images were taken at $t = 0$, $t = 5.5h$ (first visible crystallites encircled), $t = 23.5h$ (first measurable shear modulus), $t = 34.5h$, $t = 64.5h$ (plateau modulus reached) and $t = 73h$.

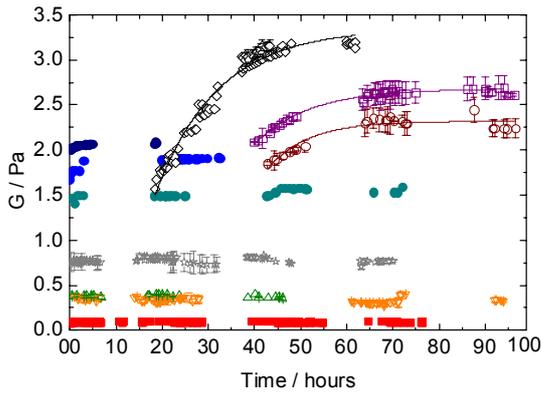

Fig. 6: (color online) Selected measurements of shear moduli of polycrystalline pure and mixed samples as a function of time. Data points group according to workin hour intervals. Filled circles: small spheres with $n$-values of $20\mu m^{-3}$ $29\mu m^{-3}$ and $30\mu m^{-3}$ from bottom to top; filled squares: large particles at $n = 2\mu m^{-3}$; open down triangles: $p = 0.2$, $n = 3\mu m^{-3}$; open up triangles: $p = 0.4$, $n = 3\mu m^{-3}$; open circles: $p = 0.9$, $n = 26\mu m^{-3}$; open squares: $p = 0.9$, $n = 29\mu m^{-3}$; open diamonds: $p = 0.9$, $n = 32\mu m^{-3}$, data from six different runs are superimposed and demonstrate the remarkably good reproducibility in time. The solid lines are fits of Eqn. (4) to the data; stars: isolated head crystal at $n = 13\mu m^{-3}$ from $p = 0.9$ and $n = 12\mu m^{-3}$.

This is shown in Fig. 6 with open symbols. For $n = 32\mu m^{-3}$ the measurements were repeated several times on the same sample. Data taken for successive runs show excellent agreement in their "kinetics". Several fit functions were tried including power laws typical for coarsening processes. Good descriptions were obtained using an exponential relation against a final value: $G(t) = G_\infty \left[1 - \exp(-t/\tau)\right]$ We obtained time constants $\tau$ of 9h, 12h and 14h for number densities of $20\mu m^{-3}$, $29\mu m^{-3}$ and $32\mu m^{-3}$, respectively. This is representative for practically all tested compositions in this composition range, which solidified within one or two days at lower $n$ and within hours at the largest $n$ investigated.

By contrast all other samples showed rather rapid crystallization. In Fig. 6 the filled circles and squares represent the pure small and large particles, respectively. Immediately after completed crystallization a few percent increase was observable, which ceased on a time scale of a few hours. It is attributed to annealing and coarsening processes of crystals and grain boundaries. Final $G$ of both pure samples increase with increased $n$ and absolute values are in good agreement with previous measurements and theoretical expectations [52]. Also the mixtures at $p < 0.7$ and $n > n_M$ readily resolidified. Fig. 6 shows for two examples at $p = 0.2$ and $p = 0.4$, that $G$ variates only little with time (open up and down triangles). We also investigated the shear modulus of a head crystal isolated at $n = 13 \mu m^{-3}$ from a larger sample at an original composition of $p = 0.9$ at an original density of $n = 12 \mu m^{-3}$. Also this sample showed a fast solidification (stars).

The absolute values of $G$ in the mixtures depend on both composition and number density. A systematic survey is under way and will be reported in a forthcoming paper. However, The preliminary data show that, within error bars, the final shear moduli of the mixtures at $p = 0.9$ and $n > n_M$ are in agreement with theoretical expectations based on the formation of a congruent solid solution [33, 1]. I.e. in contrast to the case at coexistence, here no deviation in the composition of the crystals as compared to that of the melt was detectable.

### 3.6　*Phase diagram*

A diagram displaying the equilibrated states of the samples after some two months is shown in Fig. 7. The open circles, shaded diamonds and filled squares represent fluid, coexistent and fully crystalline samples three months after the last shaking. The positions of data points represent the initially adjusted composition and number densities with small residual uncertainties in their location (for the samples where $n$ was determined from static light scattering). In Fig. 7a we use the "natural" coordinates composition and number density, which highlights the enormous suppression of crystal stability in the region $0.55 < p < 0.95$. The deviations in $n_M$ from that of the pure small spheres exceed an order of magnitude. For $p = 0.82$ the fully crystalline state is not even reached, setting a lower bound to the maximum freezing density at $n_E > 26\mu m^{-3}$.

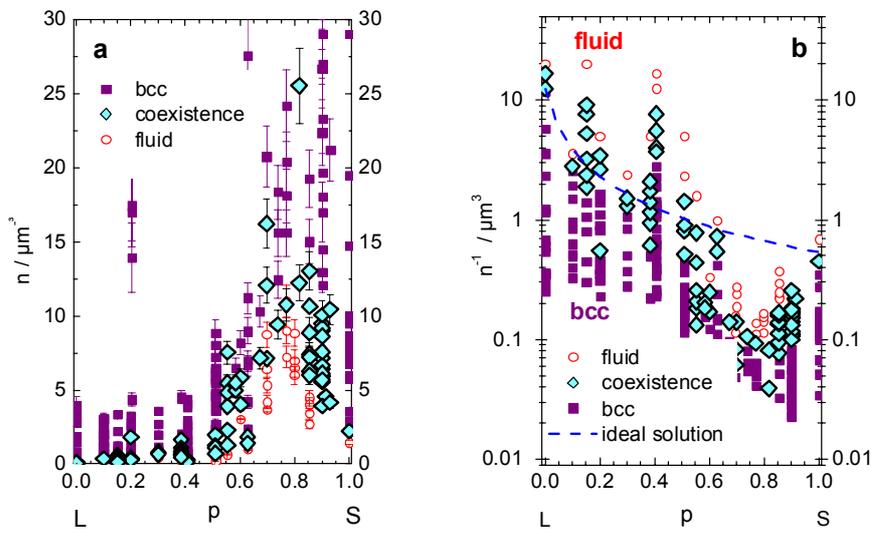

Fig. 7: (color online) Phase diagram of an exhaustively deionized charged sphere mixture with size ratio $\Gamma = 0.557$ and charge ratio $\Lambda = 0.568$ obtained from visual inspection. Left number density-composition plane, right inverse number density – composition plane. Filled squares denote bcc crystals, diamonds denote coexistence and open circles denote fluid phase. The blue dotted line is the expectation for the solidus (melting line) of spindle type phase diagram.

In Fig. 7b we show a semi-log plot of inverse number density versus composition. The fluid now is observed at the top of the graph. This representation originally suggested by [34] also allows the identification of an additional interesting feature at low $n$. An enhanced crystal stability is observed at $p \cong 0.4$. A similar, but less clearly pronounced feature is also seen at $p = 0.175$

In both plots the symbols denote the phases as observed in the presence of gravity. For the pure samples and samples with $p \leq 0.7$ solidification was rapid and occurred before any noticeable gravitational settling. Here the obtained phases are the equilibrium ones with good accuracy. Therefore the liquidus and solidus curve are formed by the freezing and melting points, $n_F(p)$ and $n_M(p)$, which are situated between the last fluid sample (circles) and the first sample at coexistence (shaded diamonds) and the last sample at coexistence and the first fully crystalline sample (filled squares), respectively. For $0.7 < p < 0.8$ we cannot exclude the existence of small amounts of residual melt in grain boundary regions. Thus here the freezing point is well defined by the symbols shown, but the transition to the fully crystalline state only marks a lower bound for the melting points. For $0.82 \leq p < 0.93$ crystal formation occurred after noticeable gravitational settling and segregation. Here the exact position of the freezing and melting points cannot be determined. The height dependent density and composition analysis however shows that both have to be shifted to larger densities and also towards slightly larger $p$. At densities denoted as fully crystalline, the crystallization kinetics again become faster. In addition, within the experimental uncertainty of the preliminary analysis of long time shear moduli, we found the formation of congruent crystals. In metal systems such behavior is observed, when the sample is rapidly quenched below the solidus. For $0.82 < p < 0.93$ we therefore estimate the position of the solidus to lie within the region marked fully crystalline in Fig. 2. For $p = 0.82$ neither freezing nor melting point can be determined. We conclude, that for $p < 0.7$ the boundaries marked by the symbols do represent the equilibrium phase boundaries, while for $0.7 \leq p < 0.93$ the equilibrium phase boundaries are located below and to the right of the marked ones. The overall shape of the phase diagram however is retained.

The representation in Fig. 7b facilitates a convenient comparison to the conventional temperature-composition or pressure-composition representation of atomic and molecular phase diagrams. The dashed blue line gives the melting line expectation for a spindle type phase diagram as derived from a linear interpolation between the $n_M$ of the pure samples. The phase diagram for our mixture deviates in both the region of strongly reduced crystal stability at $0.55 < p < 0.95$ and the sharp feature of enhanced crystal stability at $p \cong 0.4$. We therefore observe a more complex phase behavior than reported in previous studies on charged [34, 35, 36] and hard [20, 21, 22, 23, 24, 30, 31] colloidal spheres. Combinations of regions with suppressed crystal stability (lower azeotrope or eutectic) with regions of enhanced crystal stability (upper azeotrope or compound formation) are, however, observed for a number of binary metal systems [56].

*3.7*   Eutectic or azeotrope?

In metal systems one observes the formation of two mutually immiscible crystal species below a eutectic, while below a lower azeotrope a single solid solution is formed. Therefore we studied the phase behavior and morphology of selected high density samples also by microscopy in the flat cells. Fig. 8 shows a collection of images taken on a sample of $p = 0.93$ and $n = 17\mu m^{-3}$. Due to the slightly larger salt concentration in the microlife® cells this sample is still at coexistence. In Fig. 8a one observes a colorful, large, slightly facetted crystal and a collection of dark, pebble-like areas spread all over the image. Variation of the focus position revealed that the latter are restricted to the immediate vicinity of the lower glass wall, while the former appeared in the bulk on top of the latter. Fig. 8b and c compare a selected region of Fig. 8a in reflection mode (near vertical illumination) and grazing incidence mode (near horizontal illumination). In the latter case the bright green iridescence reveals the crystalline nature of the dark areas seen in reflection mode. The identical color of all scattering regions, irrespective of their in-plane orientation, shows that the crystals must be rather thin [50]. Finally, the corresponding polarization microscopic image is shown in Fig. 8d. In this mode the intensity is proportional to the thickness of the observed crystalline region and the color depends on the orientation. One recognizes that the large crystal has grown on top of the smaller dark areas remaining restricted to the vicinity of the lower cell wall and has filled gaps between these. In another sample of same composition $p = 0.93$, but for lower salt concentrations the colored crystals filled nearly the complete volume, indicating their correspondence to a high $p$ phase, while the remaining dark crystals are either pure large or at least a large sphere rich phase.

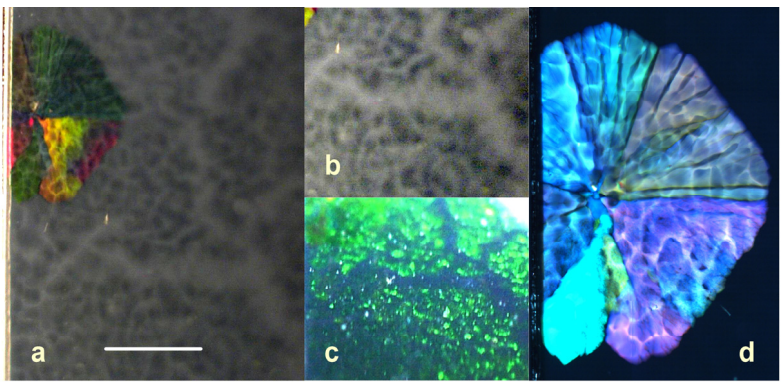

Fig. 8: (color online) Bulk and surface crystallization at coexistence above the eutectic density. a) Reflection photograph of a suspension at $p = 0.93$ and $n = 17\mu m^{-3}$. Scale bar 1mm. One observes a brightly iridescent crystal of approximately sixfold symmetric habitus at the top left. In addition a fine pattern of dark spots is visible which is spread over the complete cell wall. b) and c) comparison of a selected area in large angle reflection mode and grazing incidence mode. The formerly dark spots now show bright green Bragg scattering independent of the crystal orientation in the plane. d) close-up of the large crystal viewed between crossed polarizers. Only regions showing Bragg scattering under normal incidence illumination appear in bright colors. The white bar refers to 1mm.

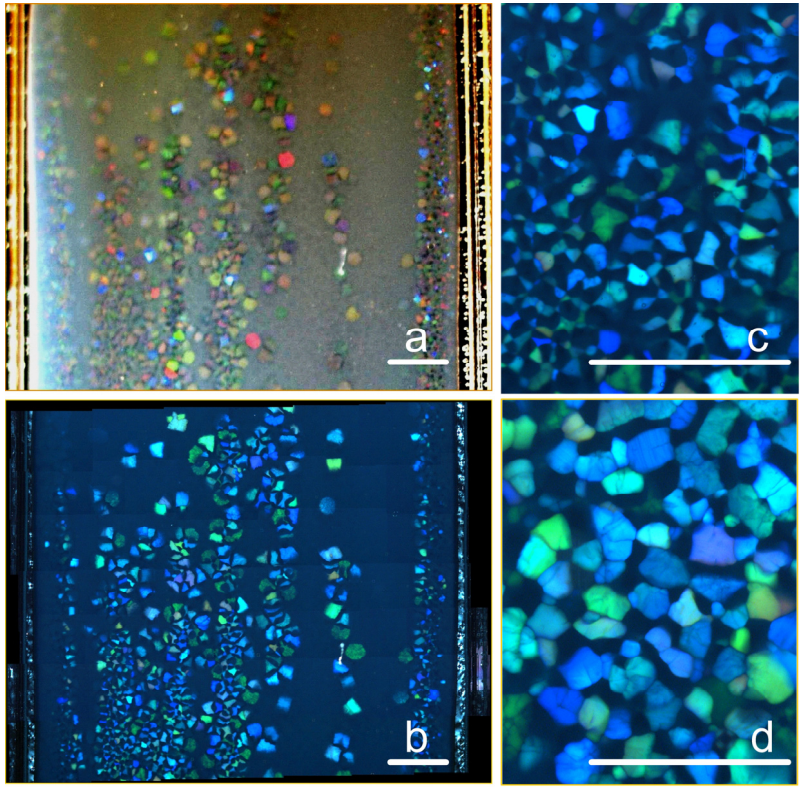

Fig. 9: (color online) Eutectic crystallization at $p = 0.9$ and $n = 43\mu m^{-3}$. All scale bars 1mm. a) reflection photograph of a partially crystallized region 4d after shaking. One notes bright Bragg reflections of crystals and dark areas. b) the same area viewed between crossed polarizers. c) close up of b) showing the facetted shape of the dark regions, which at this density grow into the bulk of the suspension. d) close-up taken of a different area 6d after shaking. Here the blue crystals have started to "overgrow" the black ones. The white bars refer to 1mm.

In Fig. 9 we show images of a suspension at $p = 0.9$ and $n = 43\mu m^{-3}$. The sample had been pre-deionized and was shaken up again. As before one recognizes dark areas at the lower cell wall and colored crystals in the reflection photograph of Fig. 9a. Fig. 9b shows the corresponding polarization microscopic image. Note, that in particular the colored species appears to be arranged in rows parallel to the formerly applied shear direction. Their different coloring, however, indicates different orientations. Also here the dark areas are crystalline. They are again spread over the lower cell wall, but now also form bulk crystals within the colored crystal rows. In the magnification of Fig. 9c one clearly sees how the dark crystals are forming facet like boundaries with the colored crystals. Two days later the colored species had still gained in volume and "overgrown" the dark one, which, however was still coexistent. Hence, at this elevated $n$ two crystal species are forming and coexisting side by side and at the same elevation under bulk conditions. Even though we did not yet analyze the composition of these two species, our finding strongly supports the classification of the stability minimum as eutectic.

*3.8* Upper azeotrope or compound?

This question was addressed by static light scattering at $p = 0.4$ and different densities above melting. In practically all samples close to melting we observed scattering patterns characteristic for a body centered cubic structure (bcc) with well indexable peaks. Only for one sample at $p = 0.4$ and $n = 2.5\mu m^{-3}$ we found additional peaks. This density is considerably above the melting density, i.e. in a region, where other authors had reported compound formation [35, 38, 39, 40]. The scattering pattern is shown in Fig. 10.

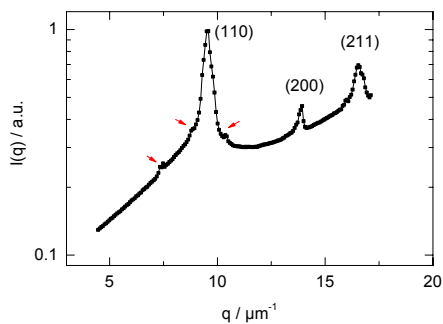

Fig. 10: (color online) Static light intensities of a sample with $p = 0.4$ and $n = 2.5\mu m^{-3}$. The scattering pattern shows a bcc crystalline structure and additional peaks indicated by arrows.

The scattering pattern shows a bcc crystalline structure and additional peaks indicated by arrows. Their position was compared to the expected positions of AB, $AB_2$, $AB_3$ $A_2B_3$ and $AB_{13}$ compounds with CsCl, NaCl, etc. but no clear coincidence was found. A somewhat better agreement was observed when the peaks were indexed as (200), (211) and (220) of a second bcc ordered crystal species of lower density. More measurements are needed to clarify this issue.

**4.    Discussion**

We have conducted a study on the phase behavior of an exhaustively deionized binary mixtures of charged spheres at a size ratio of $\Gamma = 0.557$. We observed a complex phase behaviour with a phase diagram combining composition ranges of decreased crystal stability and ranges of increased stability. From additional morphologic investigations we presented experimental evidence that the minimum of crystal stability is a eutectic. For $0.82 \leq p \leq 0.93$ we observed the gravity assisted formation of small sphere enriched head crystals of increased number density. A few points need further discussion.

First our study clearly shows that deionized charged colloidal suspensions need particular care concerning their conditioning and the measurement protocol. Our study was based on an improved batch conditioning protocol. The obtained long ranged Coulomb repulsion stabilized the crystalline phase already at particle number densities considerably lower than the glass transition density. In contrast to less deionized samples of previous studies at similar size ratio [34] we have been able to access the fluid-solid phase boundary at all compositions. Measurements of the phase diagram were started only after complete deionization monitored *via* observation of crystallite sizes [51]. This circumvented artefacts from migration within ion concentration gradients present during conditioning and allowed a detailed mapping of the coexistence region.

Second, to assess the validity of the phase transition location in a log(1/$n$) – $p$ representation we compared the time scales of gravitational settling and crystallization. In most cases crystallization is much faster than any noticeable gravitational settling. For these compositions the freezing and melting points are accurate. In the composition range of $0.82 \leq p \leq 0.93$ the gravitational separation proceeds crystallization at altered conditions. Here the location of the melting and freezing line is shifted in both composition and density with respect to the mapped-out coexistence range. The characteristic type of phase diagram, however was retained.

The influence of gravity on the distribution of phases is well known in literature [2, 15]. There phase formation proceeds gravitational settling and gravity becomes a useful means to discriminate coexisting phases. Gravity may, however also influence the crystal formation process itself. For hard spheres under μg conditions crystallization was reported at volume fractions of $\Phi = 0.62$, where under normal gravity the systems are arrested in the amorphous state under normal gravity [57]. Gravitational settling may shear off the rim of crystals [58] or control the supply limited formation of columnar crystals at the cell bottom [59]. The present case is different, as here the gravitational demixing creates the conditions necessary for crystallization in the large sphere depleted region. Moreover the gravitational enrichment of small particles at the top is conducted further by thermodynamics, when the crystallites are actually formed. So here the two processes work in the same direction.

A third point calling for attention is the shape of the observed phase diagram, which is considerably more complex than observed in previous studies at comparable Γ. We observed a broad region of decreased crystal stability at $0.55 < p < 0.95$ with the minimum located at $p$

= 0.8±0.05 at densities above $n = 26\mu m^{-3}$. Similar suppressions of the crystal phase have been reported before [34, 35], but here we could provide additional information from a morphological analysis. Observation of two coexisting crystal species at still larger densities strongly suggests the classification as a local eutectic.

An unexpected feature of our phase diagram is the pronouncedly enhanced crystal stability observed at $p = 0.4\pm0.02$ and a second stability feature at $p = 0.175\pm0.025$. We are not aware of previous reports of such an effect for colloidal suspensions. Hard spheres freeze into compounds for size ratios $\Gamma < 0.62$ with $LS_2$ yielding the most efficient packing in the range $0.62 > \Gamma > 0.48$ [20, 21, 22, 23, 24, 29, 30, 31]. Starting from the large component side and going through several compound phases the phase boundary location in $1/n$ direction continuously descends until close to $p = 1$ an eutectic is observed. At no intermediate composition the freezing density is close to or lower than that of an ideal spindle type mixture. Also for other size ratios stability maxima are neither expected nor have been reported. The maximum at $p = 0.4$ is possibly correlated with compound formation at elevated $n > n_M$. However, crystals observed for $n$ close to $n_M$ were of substitutionally ordered bcc structure. Therefore the origin of the stability features remains unresolved.

A last point concerns a subtle but nevertheless important detail of the phase behaviour observed. For our long ranged repulsive spheres we find a eutectic phase diagram type with relatively small extensions of the coexistence regions in the $p$-direction. In fact, for samples with $n > n_M$ crystallization was even congruent within experimental error. Also at coexistence after gravitational segregation only a small secondary shift of densities occurred. This behavior is at contrast with that predicted and observed in short range repulsive hard sphere systems. Hard sphere eutectics are formed already at much larger $\Gamma > 0.79$ and the crystals formed typically show only very low mutual miscibility of the two particle species [29, 30, 31]. Also metal eutectics typically form two crystal species of vastly different composition.

At this point, it is instructive to recall the connection between the phase diagram type and the qualitative concept of mutual miscibility. Spindle type phase diagrams are formed at so-called indifferent miscibility. This means that the miscibilities of the two species equal in both crystal and fluid phase. If the miscibility in the crystalline phase is reduced as compared to that in the fluid phase, the systems form azeotropes and further also eutectics. In azeotropes only a single crystal phase (typically a substitutional alloy) is formed, while in eutectics two crystal phases are formed and coexist below the eutectic temperature or above the eutectic density. Typically their composition is far off the eutectic composition and in each a pronounced enrichment with either particle species is observed. Note that this classification does not include any assumption about the absolute values of miscibility. The observed eutectic therefore does not a priori point to a generally bad miscibility of our particles. It only denotes that at a certain composition the miscibility in the crystalline state is weaker than in the corresponding fluid. This clarifies the apparent contradiction in the observation of an eutectic at good miscibility in the crystal phase. Yet a more comprehensive and more precise determination of the $p$ and $n$ resulting for the formed crystals starting from arbitrary values of these parameters is still highly desired. In this context it might be worth considering a gravity match for the particles. This could be achieved using Polystyrene particles in a $H_2O/D_2O$ mixture as solvent.

## 5.    Conclusions

We have conducted a detailed study of the phase behavior and the solidification process of an exhaustively deionized mixture of charged colloidal spheres. We have presented a complex phase diagram containing both features of suppressed and of enhanced crystal stability. We also could characterize the influence of gravity on the solidification process. Our study has shown that the phase behavior of charged sphere mixtures is far from being trivial, despite their generally good miscibility, which affords small size ratios to realize interesting phase diagram features. Our observations suggest a number of interesting further investigations including systematic studies on the occurrence of compound structures and on the elasticity of substitutional alloys. Also, the important reference experiment with gravity matched particles has not yet been performed. Finally, we intended to stimulate complementary theoretical interest in the structure formation of long range repulsive mixtures, their rich and complex phase behavior and the interesting questions raised here about mutual miscibility at moderately low site ratios and charge ratios.

## Acknowledgements

We are pleased to thank H. Löwen, E. Allahyarov, J. Horbach and P. Wette for helpful discussions. Financial support by the DFG (Pa 459/14 – 16, SFB TR6 TP D1, SPP 1120 and 1296) is gratefully acknowledged.

## References

[1]     P. M. Chaikin, J. M. di Meglio, W. Dozier, H. M. Lindsay and D. A. Weitz, *Physics of Complex and Supramolecular Fluids*, edited by S. A. Safran, N. A. Clark (Wiley-Interscience, New York, 1987, pp 65).

[2]     P. N. Pusey, *Liquids, freezing and glass transition, 51st summer school in theoretical physics, Les Houches (F) 1989*, edited by J. P. Hansen, D. Levesque, J. Zinn-Justin (Elsevier, Amsterdam, 1991, pp 763).

[3]     H. Löwen, Phys. Reports **237**, 249 (1994).

[4]     M. O. Robbins, K. Kremer and G. S. Grest, J. Chem. Phys. **88,** 3286 (1988).

[5]     P. N. Pusey and W. van Megen, Nature **320**, 340 (1986).

[6]     Y. Monovoukas and A. P. Gast, J. Colloid Interface Sci. **128**, 533 (1989).

[7]     E. B. Sirota, H. D. Ou-Yang, S. K. Sinha, P. M. Chaikin, J. D. Axe and Y. Fujii, Phys. Rev. Lett. **62**, 1524 (1989).

[8]     M. O. Robbins, K. Kremer and G. S. Grest, J. Chem. Phys. **88**, 3286 (1988).

[9]     W. van Megen, Transport Theory and Statistical Phys. **24**, 1017 (1995).

[10]    H. Löwen, J. Phys.: Condens. Matter **13**, R415 (2001).

[11]    M. E. Leunissen, C. G. Christova1, A.-P. Hynninen, C. P. Royall, A. I. Campbell, A. Imhof, M. Dijkstra1, R. van Roij and A. van Blaaderen, Nature **437**, 235 (2005).

[12]    E. V. Shevchenko, D. V. Talapin, N. A. Kotov, S. O'Brien and C. B. Murray, Nature **439**, 55 (2006).

[13]    S. Asakur and F. Oosawa, J. Chem. Phys. **22**, 1255 (1954).

[14]    A. Vrij, Pure Appl. Chem. **48**, 471 (1976).

[15]    W. C. K. Poon, J. Phys.: Condens. Matter **14,** R859 (2002).

[16]    V. J. Anderson and H. N. W. Lekkerkerker, Nature **416**, 811 (2002).

[17]    P. J. Lu, E. Zaccarelli, F. Ciulla, A. B. Schofield, F. Sciortino and D. A. Weitz, Nature **453**, 499 (2008).

[18]    A. E. Bailey, W. C. K. Poon, R. J. Christianson, A. B. Schofield, U. Gasser, V. Prasad, S. Manley, P. N. Segre, L. Cipelletti, W. V. Meyer, M. P. Doherty, S. Sankaran, A. L. Jankovsky, W. L. Shiley, J. P. Bowen, J. C. Eggers, C. Kurta, T. Lorik, Jr., P. N. Pusey and D. A. Weitz, Phys. Rev. Lett. **99**, 205701 (2007).

[19]    A. Stipp, T. Palberg and E. Bartsch, Phys. Rev. Lett. (at press 2009).

[20]    P. Bartlett, R. H. Ottewill and P. N. Pusey, J. Chem. Phys. **93**, 1299 (1990).

[21]    P. Bartlett, R. H. Ottewill and P. N. Pusey, Phys. Rev. Lett. **68**, 3801 (1992).

[22]    P. N. Pusey, W. C. K. Poon, S. M. Ilett and P. Bartlett, J. Phys.: Condens. Matter **6,** A29 (1996).

[23]    A. B. Schofield, P. N. Pusey and P. Radcliffe, Phys. Rev. E **72**, 031407 (2005).

[24]    S. Underwood, W. van Megen and P. N. Pusey, Physica A **221**, 438 (1995).

[25]    S. Hachisu, Y. Kobayashi and A. Kose, J. Colloid Interface Sci. **42**, 342 (1973).

[26]    S. Hachisu and S. Yoshimura, Nature **283**, 188 (1980).

[27]    A. Kose, M. Ozaki, K. Takano, Y. Kobayashi and S. Hachisu, J. Colloid Interface Sci. **44**, 330 (1973).

[28]    A. Kose and S. Hachisu, J. Colloid Interface Sci. **55**, 487 (1976).

[29]    W. J. Hunt, R. Jardine and P. Bartlett, Phys. Rev. E **62**, 900 (2000).

[30]    X. Cottin and P. A. Monson, J. Chem. Phys. **102**, 3345 (1995).

[31]    S. Punnathannam and P. A. Monson, J. Chem. Phys. **125**, 024508 (2006).

[32]    H. Xu and M. Baus, J. Phys. Chem. **118,** 5045 (2003).

[33]    H. M. Lindsay and P. M. Chaikin, J. Chem. Phys. **76**, 3774 (1982).

[34]    A. Meller and J. Stavans, Phys. Rev. Lett. **68**, 3646 (1992).

[35]    T. Okubo and H. Fujita, Colloid and Polymer Science **274**, 368 (1996).

[36]    P. Wette, H. J. Schöpe and T. Palberg, J. Chem. Phys. **122**, 144901 1-8 (2005).

[37]    P. Wette, H. J. Schöpe, R. Biehl and T. Palberg, J. Chem. Phys. **114**, 7556 (2001).

[38]    T. Okubo and H. Ishiki, Coll. Polym. Sci. **279**, 571 (2001).

[39]    T. Okubo J. Chem. Phys. **93**, 11 (1990).

[40]    L. Liu, S. Xu, J. Liu and Z. Sun, J. Colloid Interface Sci. **326**, 261 (2008).

[41]    P. D. Kaplan, J. L. Rouke, A. G. Yodh and D. J. Pine, Phys. Rev. Lett. **72**, 582 (1994).

[42]    J. B. Perrin, Annales de Chimie et de Physique **18**, 1-114 (1909).

[43]    D. M. E. Thiesweesie, A. P. Phillipse, G. Nägele, B. Mandl and R. Klein, J. Colloid Interface Sci. **176**, 43 (1995).

[44]    S. Buzzaccaro, A. Tripodi, R. Rusconi, D. Vigolo and R. Piazza, J. Phys.: Condens. Matter **20**, 494219 (2008).

[45]    T. Palberg and M. Würth, Phys. Rev. Lett. **72**, 786 (1994).

[46]    M. Medebach, R. Chuliá Jordán, H. Reiber, H. J. Schöpe, R. Biehl, M. Evers, D. Hessinger, J. Olah, T. Palberg, E. Schönberger and P. Wette, J. Chem. Phys. **123**, 104903 (2005).

[47]    P. Wette and H. J. Schöpe, Phys. Rev. E **75**, 051405 1-12 (2007).

[48]  J. Liu, A. Stipp and T. Palberg, Prog. Colloid Polym. Sci. **118**, 91 (2001).

[49]  T. Palberg, J. Phys.: Condens. Matter **11**, R323 (1999).

[50]  W. Loose and B. J. Ackerson, J. Chem. Phys. **101**, 7211 (1994).

[51]  J. Liu, H. J. Schöpe and T. Palberg, J. Chem. Phys. **116**, 5901 (2002) and *ibid.* **123**, 169901 (2005) E.

[52]  L. K. Cotter and N. A. Clark, J. Chem Phys. **86**, 6616 (1987).

[53]  P. Wette, H. J. Schöpe and T. Palberg, Colloid Surf. *A* **222**, 311 (2003).

[54]  P. Pieranski, E. Dubois-Violette, F. Rothen and L. Strzlecki, J. PhysiqueI (Paris) **41**, 369 (1980).

[55]  H. M. Lindsay and P. M. Chaikin, *J. Chem. Phys.* **76**, 3774 (1982).

[56]  T. B. Massalski and H. Okamoto (EDS.) *Binary alloy phase diagrams* (ASM International, Materials Park, Ohio, 2$^{nd}$ Ed. 1990).

[57]  J. Zhu, M. Li, R. Rogers, W. Meyer, R. H. Ottewill, STS-73 Space Shuttle Crew, W. B. Russel, and P. M. Chaikin, Nature **387**, 883 (1997).

[58]  Y. He, B. Olivier and B. J. Ackerson, Langmuir **13**, 1408 (1997).

[59]  B. J. Ackerson, S. E. Paulin, B. Johnson W. van Megen and S. Underwood, *Phys. Rev. E* **59**, 6903 (1999)**.**